\documentclass[journal=acsnanoletter,
manuscript=article]{achemso}
\usepackage{chemformula} 
\usepackage[T1]{fontenc} 
\usepackage{amsmath}
\usepackage{siunitx}
\usepackage{mathtools}
\usepackage{bm}       
\usepackage{amsfonts} 
\usepackage{amssymb}  

\usepackage{bm}
\usepackage{braket}

\author{D. Gill}
\affiliation[BigPharma]
{Max-Born-Institute for Non-linear Optics and Short Pulse Spectroscopy, Max-Born Strasse 2A, 12489 Berlin, Germany}
\author{S. Shallcross}
\affiliation[BigPharma]
{Max-Born-Institute for Non-linear Optics and Short Pulse Spectroscopy, Max-Born Strasse 2A, 12489 Berlin, Germany}
\author{W. Chen}
\affiliation[BigPharma]
{Max-Planck-Institut fur Mikrostrukturphysik Weinberg 2, D-06120 Halle, Germany\\
Max-Born-Institute for Non-linear Optics and Short Pulse Spectroscopy, Max-Born Strasse 2A, 12489 Berlin, Germany}
\author{J. K. Dewhurst}
\affiliation[BigPharma]
{Max-Planck-Institut fur Mikrostrukturphysik Weinberg 2, D-06120 Halle, Germany}
\author{S. Sharma}
\email{sharma@mbi-berlin.de}
\affiliation[BigPharma]
{Max-Born-Institute for Non-linear Optics and Short Pulse Spectroscopy, Max-Born Strasse 2A, 12489 Berlin, Germany \\
Institute for theoretical solid-state physics, Freie Universit\"at Berlin, Arnimallee 14, 14195 Berlin, Germany}

\title[An \textsf{achemso} demo]
  {Coupled femto- excitons, free carriers and light}

\begin{document}

\begin{abstract}
Non-equilibrium quantum matter generated by ultrafast laser light opens new pathways in fundamental condensed matter physics, as well as offering rich control possibilities in "tailoring matter by light". Here we explore the coupling between free carriers and excitons mediated by femtosecond scale laser pulses. Employing monolayer WSe$_2$ and an {\it ab-initio} treatment of pump-probe spectroscopy we find that, counter-intuitively, laser light resonant with the exciton can generate massive enhancement of the early time free carrier population. This exhibits complex dynamical correlation to the excitons, with an oscillatory coupling between free carrier population and exciton peak height that persists. Our results both unveil "femto-excitons" as possessing a rich femtosecond dynamics as well as, we argue, allowing tailoring of early time light-matter interaction via laser pulse design to control simultaneously excitonic and free carrier physics at ultrafast times.
\end{abstract}

\section{Introduction}

Traditionally light has been used as a probe for material properties, however in the present era of atto- and femtosecond lasers the role of light has become two-fold. Pump laser light generates a non-equilibrium state of matter, and probe laser light then explores the nature of this new quantum state of matter~\cite{nie2014,sie2017,mai2014,pogna2016,shang2015,lucchini2021,sim2013}. This has led to emergence of new fields of physics such as femtomagnetism~\cite{kirilyuk2010ultrafast,gonccalves2016dual}, lightwave control over spintronics~\cite{vzutic2004,ouyang2020}, and ultrafast valleytronics~\cite{kobayashi2023} to name but a few examples. Pump laser photons impart their energy to matter creating an out-of-equilibrium quantum state, the probing of which is highly complex; photons (of pump as well as probe light), electrons and nuclei \emph{dynamically couple} with each other both during and after excitation. For example, light interacts with spins via spin-orbit coupling, which is, in principle, a material-specific property. However, the pump light pulse generates a non-equilibrium state of excited charge, and these excited electrons act as a new state of matter and hence feel a spin-orbit coupling different from the original material, making the coupling dynamical\cite{krieger2015laser,liao2023state,lopez2013graphene} and hence also a property of the light and not just the material. 

In semiconductors, one such quasi-particle that is created and probed by light is the exciton; light induces electrons to excite to the conduction band (CB) leaving behind a hole, and this hole and the excited electron then form a bound electron-hole pair called exciton. Such excitons form a localized band in the gap, dramatically changing the electronic structure of the material. Probing the dynamics of such excitons upon light wave pumping requires treatment of pump light, electrons, excitons, and probe light all on the same footing. A method capable of doing so at present is the Kohn-Sham Proca (KSP)\cite{dewhurst2025kohn} approach to time-dependent density functional theory (TD-DFT)\cite{runge1984}, which is a fully {\it ab-initio} pump-probe spectroscopic approach. 

In the present work, using KSP, we unravel the complex light-wave-exciton dynamics in a monolayer (ML) of transition metal dichalcogenide (TMDC), WSe$_2$.  We pump the material and probe it at various time delays, allowing both pump and probe pulses to transiently create and destroy excitons, which also dynamically interact with free carriers. We show that this dynamics is highly complex; counter intuitively, pumping the material at the exciton energy level leads to a dramatically enhanced creation of free carriers in the CB as compared to pumping in the CB. The excitonic response further shows an intricate interplay of interacting excitons and free-carriers, mediated by the pump pulse, leading to persistent  oscillations in free carrier density in step with exciton binding energy and intensity.

\section{Results}

\begin{figure}[t!]
        \centering
        \includegraphics[width=0.85\columnwidth,clip]{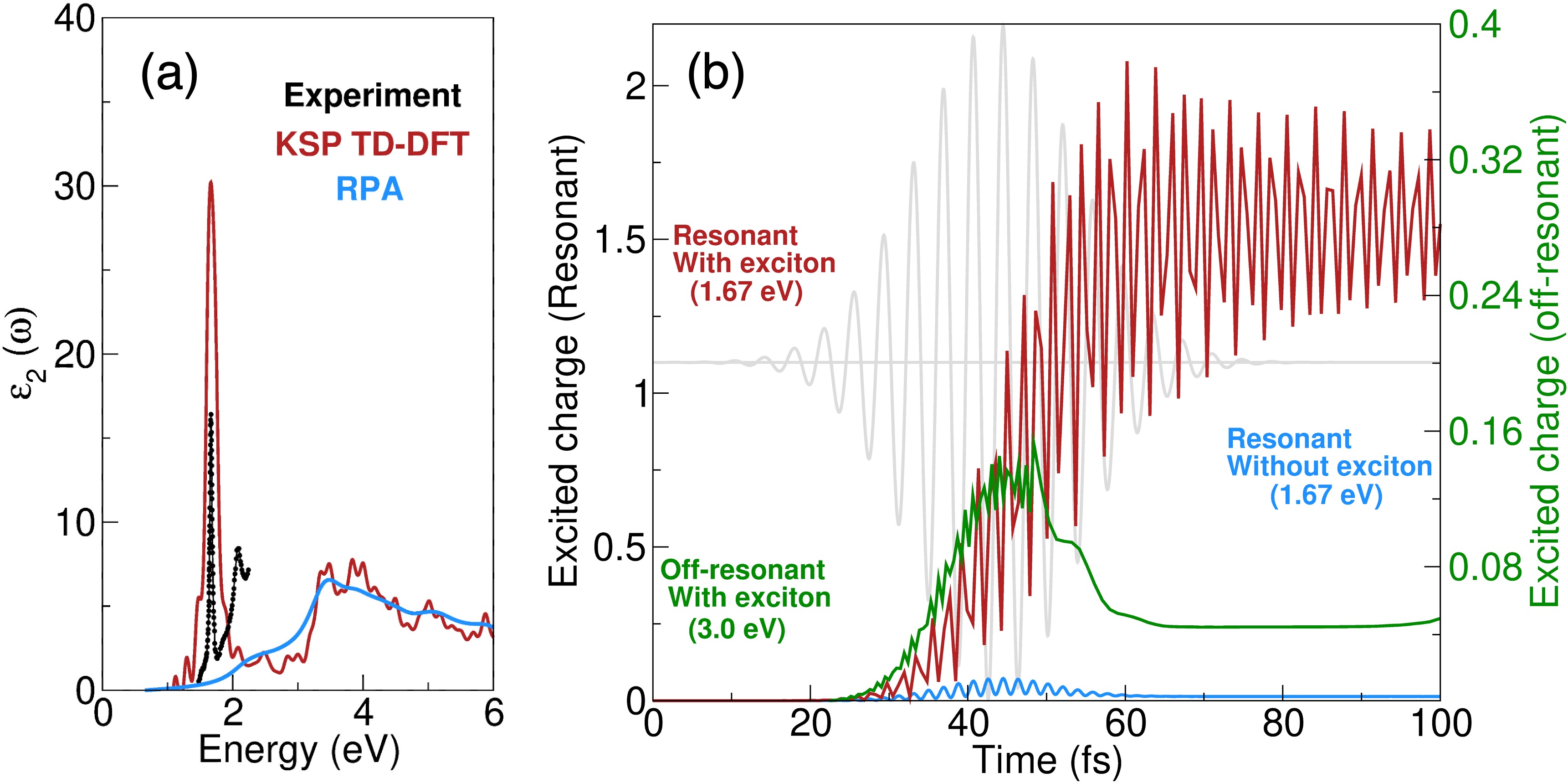}
        \caption{\emph{Influence of exciton dynamics on free carriers}; (a) Absorption spectra in  the presence and absence of excitons, for comparison experimental data\cite{aslan2018strain} is also shown. (b) Dynamics of the excited charge in presence (red) and absence (blue) of the excitons for resonant (at $\omega$=1.67 eV) and in presence of exciton dynamics for off-resonant (at $\omega$=3.0 eV) pumping (green). Vector potential of the resonant pump pulse is shown in grey.  Resonant pumping strongly couples photons, excitons and free carriers to (i) dramatically enhance the free carrier population and (ii) generate a persistent oscillatory coupling between the exciton and free-carrier populations.}
        \label{fig1}
\end{figure}

In order to unravel the complex interplay of light, excitons, and fermions, we study the dynamics of these in an example system of a ML-WSe$_2$. We drive this TMDC out of equilibrium by selective laser pumping in two different ways; (a) resonant pumping -- the material is pumped with laser light of central frequency equal to the excitonic frequency (1.67 eV) leading to exciton creation and (b) off-resonant pumping -- here the material is pumped with laser light of central frequency (3.0 eV) above the band gap. Since ML-WSe$_2$ is an insulator with a direct gap of 2.2~eV (indirect gap of 2.12~eV)\cite{zhang2015probing} and exciton binding energy~0.40 eV\cite{aslan2021excitons}, a resonant pump pulse does not allow for any significant direct creation of excited charge in the conduction band i.e., free carriers. In contrast, off-resonant pumping allows for direct creation of free carriers, some of which will form excitons. In order to ensure (a) and (b), the pump pulse is chosen to have fluence (1.70 mJ/cm$^2$), which allows significant excitation while remaining in the regime in which single-photon processes dominate, and duration (25 fs) such that the selected central frequency dominates (for the frequency spectrum of the pulses see Fig. 1 of SI). Following the excitation with these two types of pulses, we then explore the dynamics of free carriers as well as excitons. Our theoretical method of choice is the fully {\it ab-initio} Kohn-Sham-Proca (KSP) method\cite{dewhurst2025kohn,sharma2024direct} within TD-DFT (for details of the method and computations, see the Method section). This method is capable of treating the dynamics of excitons and free carriers on the same footing while they both interact with the laser pump and probe pulses.  

\emph{Free carrier dynamics}: Employing the KSP scheme we calculate the static excitonic spectra\cite{sharma2024direct} of WSe$_2$, see Fig.~\ref{fig1}(a); the exciton peak in the response function calculated using KSP (red curve) agrees well with the experimental (black curve) absorption spectra~\cite{aslan2018strain}; in close agreement with experiments we find exciton binding energy to be 0.44eV. As expected, excitons are missing from the RPA absorption spectra i.e. when the TD-DFT KS equations are solved in the absence of the Proca equation (blue curve). Having captured the correct static response of the material to light, we now follow the dynamics of the material upon resonant and off-resonant light wave pumping. The dynamics of free carries (the number of electrons in the conduction band) is shown in Fig.~\ref{fig1}(b); in striking contrast to off-resonant pumping (green), in the case of resonant pumping (red), there is an enhancement in free carrier density (by a factor of $\sim$10). There can be only two causes of such free carrier excitation for light pumping at a sub-gap frequency: (i) two photon processes or (ii) the creation of free carriers via light-wave interaction with excitons.
The former can be ruled out by inspection of the excited charge density in absence of excitons (i.e. by solving KS equations alone) and, as can be seen in Fig.~\ref{fig1}(b), in this case (blue) the excited charge falls to zero. From these results the picture that emerges is that resonant laser pulse does not just create excitons, but also dynamically interacts with these excitons to generate extra free carriers.


\begin{figure}[t!]
        \centering
        \includegraphics[width=0.85\columnwidth,clip]{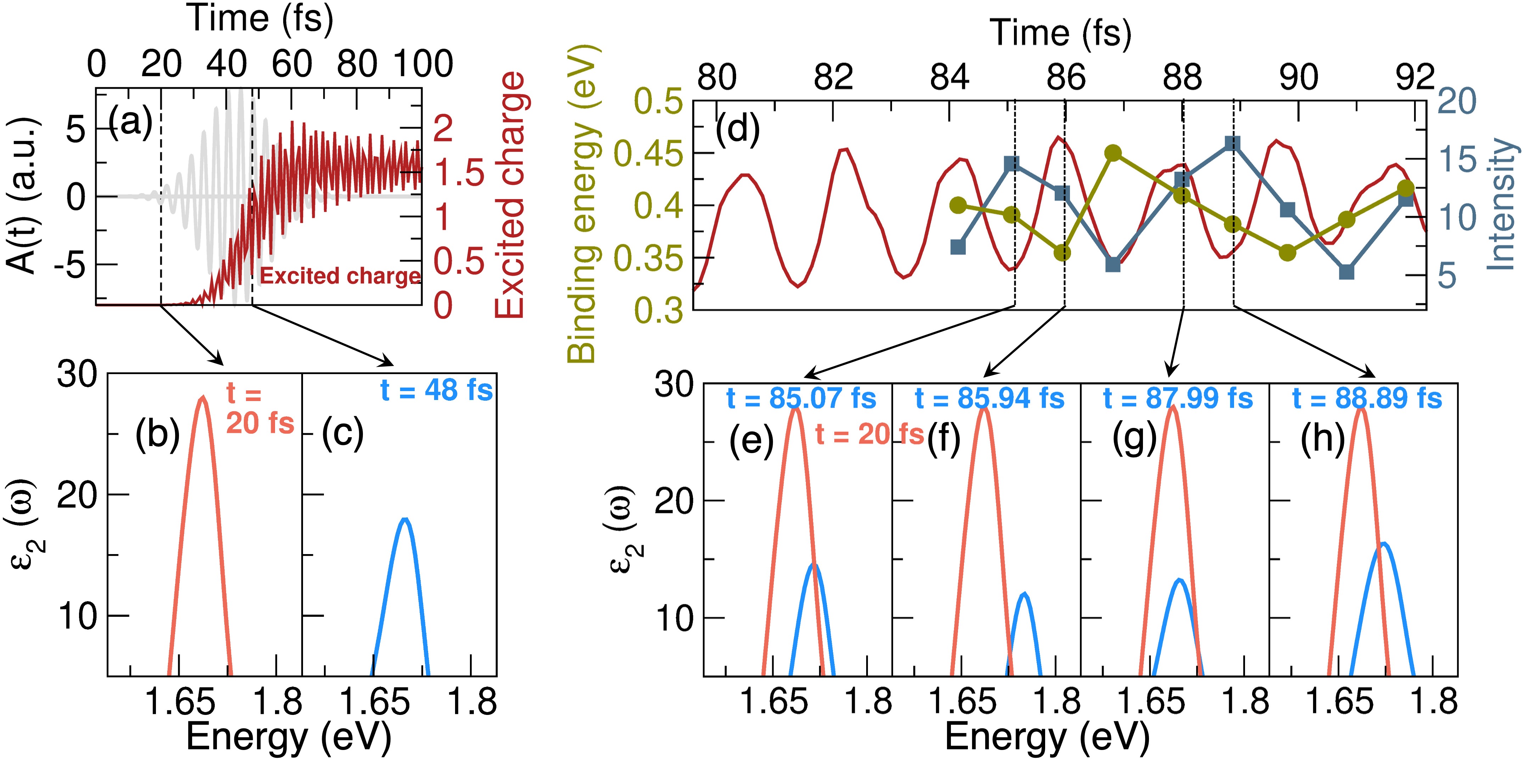}
        \caption{\textit{Strong coupling of photons, electrons, and excitons}: (a) dynamics of resonantly pumped free carriers (red) in the conduction band (vector potential of the pump pulse is shown in grey, see computation details for pulse parameters). Dotted black lines indicate times at which material is probed and the  corresponding absorption spectra is shown in panels (b,c). (d) Dynamics of exciton binding energy (green) and peak intensity (blue) together with the dynamics of excited free carriers (red) in a time window 80-90 ~fs. Dotted black lines indicate times at which the  absorption spectra is shown in panels (e-h).}
        \label{fig2}
\end{figure}

\emph{Exciton dynamics}:  We now follow the dynamics of these excitons (see Fig.~\ref{fig2})  and for this we will use the pump-probe spectroscopic method, wherein we probe the excitonic response of ML-WSe$_2$ at various time delays after pumping, as would be done in any potential spectroscopic experiment (for details and a schematic of this process see Fig.~\ref{fig4} in method section). Upon resonant pumping, excitons precede free carriers; by probing WSe$_2$ during the rising edge of the pump pulse and before any free carrier excitation (see first dotted lines in Fig.~\ref{fig2}(a)) we see an excitonic peak in the response, indicating exciton formation. However, as the pump pulse reaches its maximum amplitude, there is a significant bleaching of the excitonic response i.e. decrease in excitonic peak intensity (see Fig.~\ref{fig2}(c)). This decrease in peak intensity is an indicator of destruction of excitons due to laser pumping and this happens at the same time as there is an increase in free carrier density, further cementing the fact that excitons and free carriers show a correlated dynamics in ultrafast laser pumping. 

\begin{figure}[ht!]
        \centering
        \includegraphics[width=0.95\columnwidth,clip]{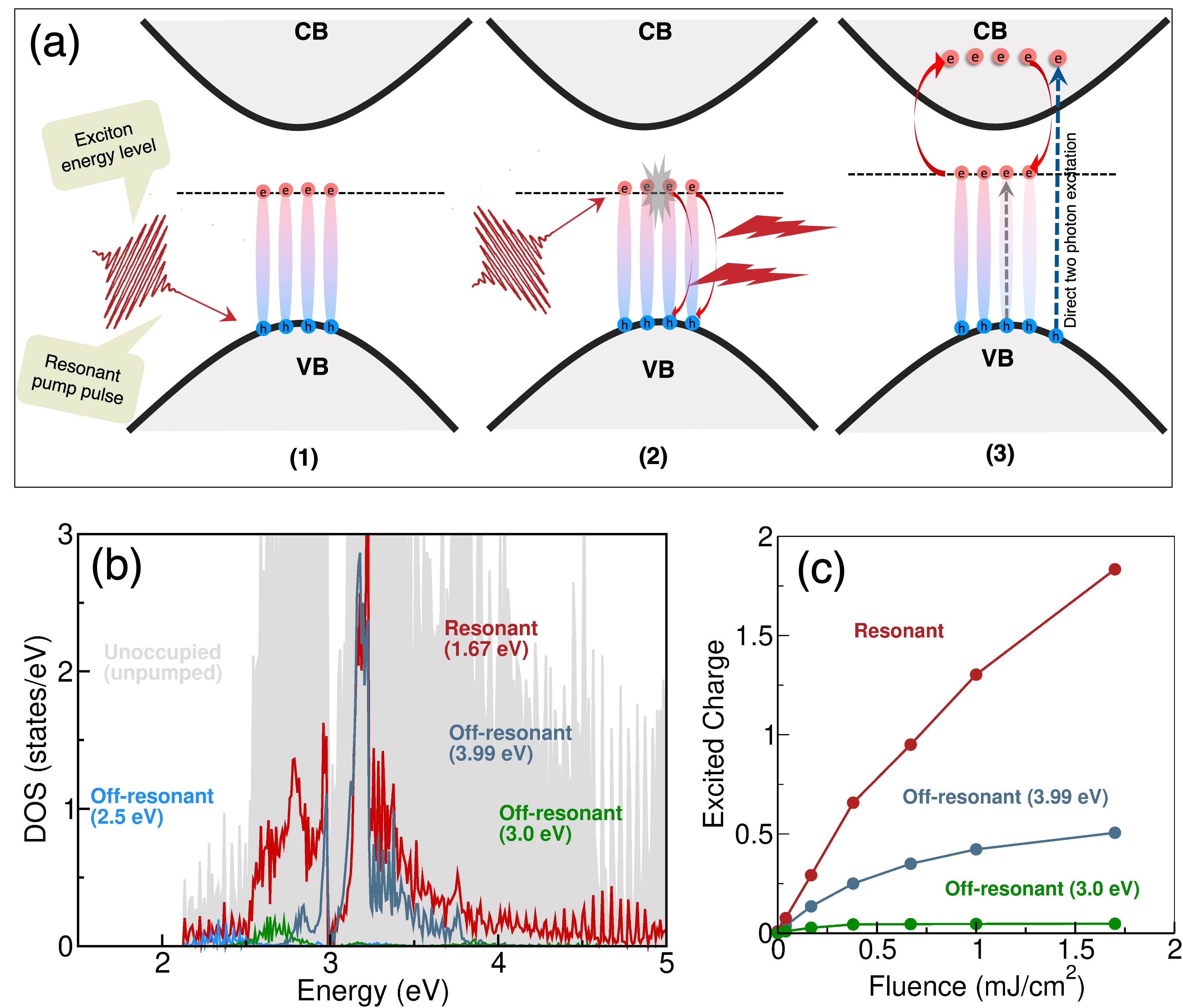}
        \caption{(a) Schematic illustration of exciton mediated free carrier excitation: (1) the laser pump creates excitons by resonant excitation; (2) exciton-light and exciton-exciton interaction drives exciton recombination and photon emission; (3) these photons then lead to excitation of charge into the conduction band. (b) The unoccupied ground state density of states (DOS), illustrated in grey, reveals a low density of available states at the CB edge with a much increased density of available states at higher energies within the CB. For off-resonant pumping, direct excitation to the conduction band edge quickly saturates due to Pauli blocking, with the transient DOS of the excited charge shown in light-blue, green and dark-blue for three different frequencies of the pump pulse. In contrast, resonant pumping excites particles into a region with a high density of available states, facilitating significant generation of free carriers (shown in red). (c) Excited charge as a function of fluence for resonant (red) and off-resonant (green and blue) pumping. In the former case an increase with the fluence is observed, while in the latter cases saturation occurs at very low fluence.}
        \label{fig3}
\end{figure}

This overall bleaching of the excitonic peak upon light wave pumping is in agreement with past experimental observations\cite{pogna2016,aharon2016new,baldini2018clocking,fluegel1997exciton}. However, a closer inspection reveals that the dynamics of the excitonic peak intensity (during and after pumping) is more complex than a steady bleaching. By probing the material at various time delays, we find that the excitonic peak intensity as a function of time is oscillatory (see blue curve Fig.~\ref{fig2} (d)); this is an indicator of dynamical formation and destruction of excitons under the impact of a pump pulse. 
Such oscillatory response of the excitonic peak intensity due to dynamical destruction of excitons has recently been seen in experiments
~\cite{timmer2024ultrafast,souri2024ultrafast} in the context of phonon mediated inter-valley coupling\cite{dong2021direct}, occurring on picosecond time scales. The ultrafast time scale of the oscillation seen here implies a mechanism involving only electrons and light. To explore this, we note that for most of the time delays the reduction in exciton peak intensity is correlated with an increase in free carrier density (red and blue curves Fig.~\ref{fig2} (d) and absorption spectra Fig.~\ref{fig2} (e)-(h)) and vice versa. A similar process was experimentally unveiled in carbon-nano-tubes where a decrease in exciton peak was correlated with an increase in population of second excitonic state\cite{ma2005femtosecond}, instead of CB as in present case.
In the present case binding energy of the excitons also shows in-step oscillations (see green curve in Fig.~\ref{fig2} (d)), indicating the role played by screening in breaking of the excitons\cite{smejkal2020}. We note that the changes in the KS band gap\cite{cunningham2017photoinduced,pogna2016,chernikov2015population,liu2019direct} during this early time dynamics is very small. 
A picture that now emerges is thus of a dynamical system of correlated excitons, free carriers and light: (1) upon resonant pumping excitons are first created; (2) then upon interaction with the pump laser pulse and with each other, some of these excitons are destroyed; (3) this leads to an increase in free carrier density; some of these free carriers contribute towards screening of excitons while at the same time some of these free carriers form new excitons (for schematic see Fig.~\ref{fig3} (a)). 
Given that our system is isolated and decoupled from phonons (and any other energy loss mechanism), the cycle of excitation and de-excitation of free carriers continues.

\emph{Availability of states and pulse frequency}: From these results it is clear that the free-carriers and excitons show early time correlated dynamics mediated by the pump laser pulse. We can further investigate the underlying microscopic physics of this dynamics, and role of pulse frequency, by inspecting the transient electronic structure of ML-WSe$_2$. Upon resonant pumping dissociation of excitons generates free carriers that  are markedly different from those generated by off-resonant pumping. The former entails a two-step process involving both the creation and destruction of excitons, with the transient free carrier population found at twice the exciton energy, as may be seen in the transient excited density of states (DOS) Fig.~\ref{fig3}(b) red curve. Off-resonant pumping, on the other hand, generates free carriers by direct optical excitation into the conduction band. In Fig.~\ref{fig3}(b) such results are shown for three different excitation frequencies (light-blue, green, and dark--blue curves);-- the density of excited free carriers in this case of off-resonant pumping is determined by the density of available states (shown as a grey background). However, the amount of excited charge for the off-resonant pumping is always lower than that found in the resonant case, where exciton dissociation also contributes to free carrier generation.


\emph{Laser pulse fluence}: These results indicate the important role played by the pump pulse parameter, namely the central frequency, in mediating free-carrier exciton-correlated dynamics. When changing this central frequency the incident fluence of the pump pulse was kept constant at 1.7 mJ/cm$^2$, i.e. the pulse is intense enough to break (and not just create) excitons. Since breaking of the excitons is instrumental in detecting the correlated dynamics of free carriers and excitons, we expect non-linearity in free carrier density upon varying the fluence. Indeed we see that after an initial linear increase in free carrier density with increasing fluence the curve starts to turn for resonant pumping (red curve Fig.~\ref{fig3}(c)) and saturates for off-resonant pumping (green and blue curves Fig.~\ref{fig3}(c)). The reason for the saturation for the off-resonant pumping is Pauli blocking\cite{zhu2018breaking,iwasaki2023electronic,
amo2007pauli} and inability of off-resonant pulse in breaking excitons at ultrafat times, while in the resonant case either the process of exciton dissociation via light slows down and/or exciton-exciton dissociation is Auger like\cite{sun2014observation} leading to the turning. 
Such breaking or recombination of excitons after pumping and an associated linear dependence upon pump fluence (i.e. increased dissociation with fluence) has been observed experimentally in TMDCs\cite{wang2015surface,sun2014observation,
kumar2014exciton,kar2015probing}. These works, that focus on later picosecond times in the regime of excitonic scattering and relaxation, have also revealed a complex intertwined dynamics of excitons, electrons and holes. In the present work, by studying sub-cycle dynamics, we show that such complex correlated dynamics occurs even at early femtosecond times in which there is correlation also with the pump light itself.

\section{Discussion}

At the picosecond times at which a laser excited state returns to equilibrium, recent experiments report a coupling of excitons, free carriers, and lattice excitations. Here we have found that already in the sub-cycle structure of the laser pulse excitons and free carriers are intimately coupled. This  underpins a very rich early time dynamics in which the populations of free carriers and excitons exhibit strong dependence on each other. A notably counter intuitive example of this being that laser pumping resonant with the exciton generates a free carrier population significantly greater than can be obtained by laser pumping directly into the conduction band, an effect driven by a combination of light induced dissociation of excitons and Pauli blocking.

This fundamentally intertwined nature of light, free carriers, and excitons in femtosecond non-equilibrium dynamics is dramatically illustrated by an in-step oscillation of free carrier and exciton populations and should be observable in experiments with good time resolution.

Our results thus point towards a rich control over exciton dynamics via free carrier excitations and, vice versa, upon free carriers via exciton pumping mediated by the laser pump pulse. Furthermore, since availability of excited states plays a crucial role in the density of free carriers as well as excitonic properties (such as binding energy and intensity), laser pulse design and manipulation of the density of states -- both now raised to an art forms in modern condensed matter research -- promise rich possibilities for tailoring non-equilibrium excitonic states by light, i.e. femto-excitonics.

\section{Methodology}

\emph{Kohn-Sham-Proca method:} TD-DFT \cite{runge1984,sharma2014} rigorously maps the computationally intractable problem of interacting electrons to a Kohn-Sham (KS) system of non-interacting electrons in an effective potential. The time-dependent KS equation is:

\begin{align}  
\begin{split}
i \frac{\partial \psi_{j}({\bf r},t)}{\partial t} =
\Bigg[
\frac{1}{2}\Big(-i{\nabla}&-\frac{1}{c}\big({\bf A}(t)+{\bf A}_{\rm xc}(t)\big)\Big)^2 
+ v_{s}({\bf r},t) \Bigg]
\psi_{j}({\bf r},t),
\end{split}
\label{e:TDKS}
\end{align}
where $\psi_j$ are the KS orbital and the effective KS potential $v_{s}({\bf r},t) = v({\bf r},t)+v_{\rm H}({\bf r},t)+v_{\rm xc}({\bf r},t)$ consists of the external potential $v$, the classical electrostatic Hartree potential $v_{\rm H}$ and the exchange-correlation (XC) potential $v_{\rm xc}$. The vector potential ${\bf A}(t)$ represents the applied laser field within the dipole approximation (i.e., the spatial dependence of the vector potential is absent) and ${\bf A}_{\rm xc}(t)$  the XC vector potential. This is generated by coupling the KS equation (Eq. \ref{e:TDKS}) to the Proca equation

\begin{align}
a_2\frac{\partial^2}{\partial t^2}{\bf A}_{\rm xc}(t)+a_0{\bf A}_{\rm xc}(t)= 4 \pi {\bf J}(t),
\label{e:proca}
\end{align}
here the gauge invariant current, {\bf J}, is obtained by integrating the microscopic current density {\bf j}, which is given by:

\begin{align}
\begin{split}
 {\bf j}({\bf r},t)=
 &{\rm Im}\sum_j^{\rm occ}\psi_j ({\bf r},t) ^*{\nabla}\psi_j({\bf r},t) -\frac{1}{c}\big({\bf A}(t) + {\bf A}_{\rm xc}(t)\big)\rho({\bf r},t). 
\end{split}
\label{e:j}
\end{align}
Thus Eqs. \ref{e:TDKS} and \ref{e:proca} are coupled and are simultaneously solved (further details of this method see Ref.~\cite{dewhurst2025kohn}). The ${\bf A}_{\rm xc}(t)$ generated in this manner then allows one to treat the light-induced dynamics of excitons and free carriers at the same footing. 

\begin{figure}[ht!]
        \centering
        \includegraphics[width=0.8\columnwidth,clip]{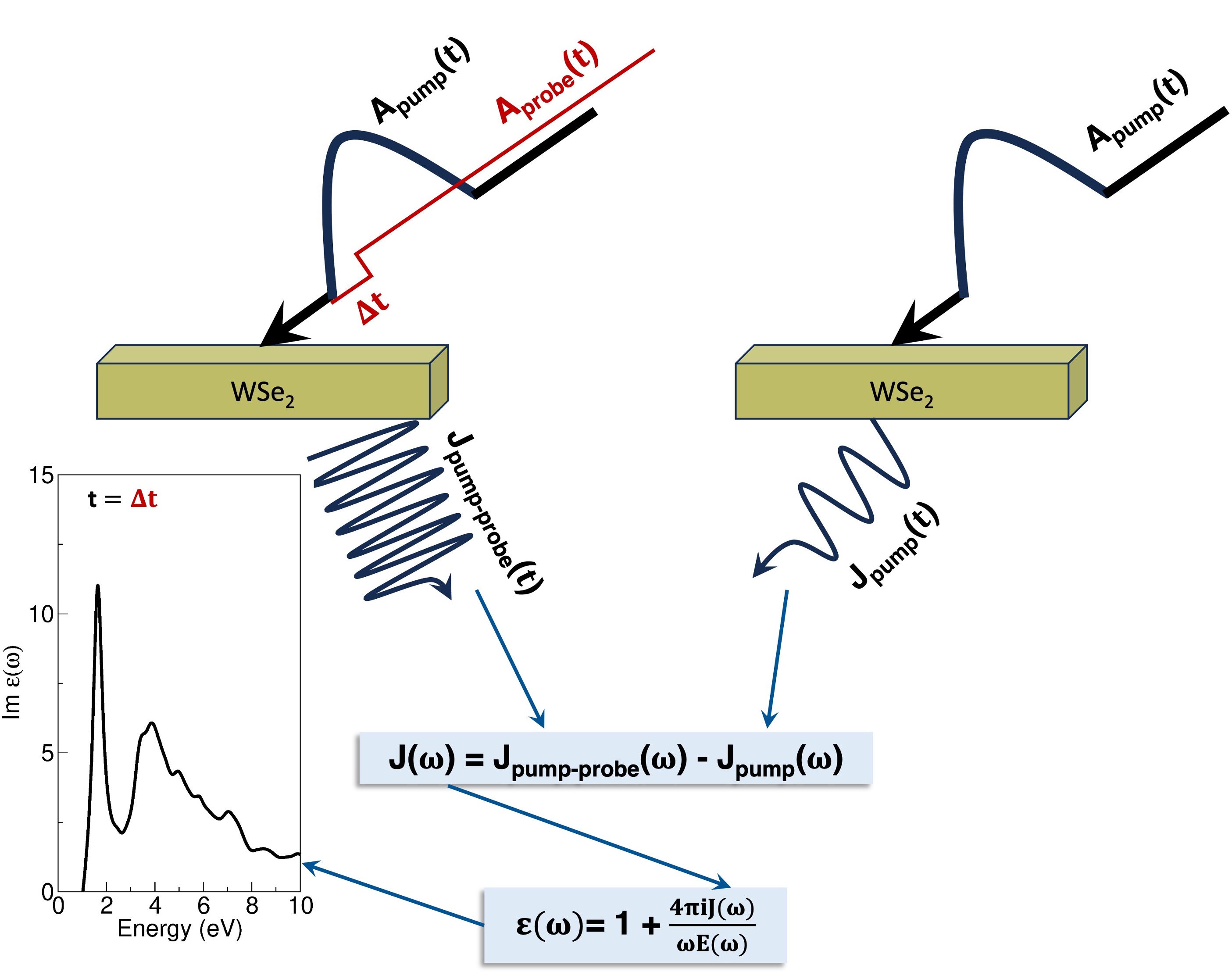}
        \caption{\textit{Schematic illustrating the calculation of transient excitonic response using pump-probe approach}. The charge current in the frequency domain J($\omega$), which is obtained by taking the difference of the current J$_{\rm{pump-probe}}$($\omega$) (current obtained when the system is exposed to pump-probe pule) and J$_{\rm{pump}}$($\omega$) (the current obtained when the system is exposed to pump pulse), is used to calculate the dielectric function containing the information of the excitonic physics at time t = $\Delta$t.}
        \label{fig4}
\end{figure}

The transient excitonic response is determined using the pump-probe method, as illustrated in Figure~\ref{fig4}. In this approach, the system is pumped with a laser pulse (with vector potential A$_{\rm{pump}}$(t)) and then probed, after a time delay of $\Delta t$, with a very weak probe pulse (with vector potential A$_{\rm{probe}}$(t)). This yields two kinds of currents; J$_{\rm{pump}}$(t) and J$_{\rm{pump-probe}}(t,\Delta t)$. The difference between the Fourier transform of these currents ( J($\omega, \Delta t$) = J$_{\rm{pump-probe}}$($\omega, \Delta t$) -J$_{\rm{pump}}$($\omega$)) is then used to generate the response function:

\begin{align}
\varepsilon(\omega,\Delta t) = \frac{4\pi i\rm{J}(\omega,\Delta t)}{\omega \rm{E}(\omega)},
\label{eps}
\end{align}
where $\varepsilon$($\omega,\Delta t$) in the dielectric function and E($\omega$) is the Fourier transform of the Electric field of the probe laser pulse, E(t) = -1/c($\delta A_{\rm{probe}} (t)/\delta t)$.  

\emph{Computational details}: In our work, we used the adiabatic local density approximation for the exchange-correlation potential and all calculations were performed using the highly accurate full potential linearized augmented-plane-wave method, as implemented in the ELK code (version elk-10.3.12)\cite{elk}. The ML-WSe$_2$ was modelled using a hexagonal unit cell, with in-plane lattice vectors\cite{schutte1987crystal}: $a$ = [1.658, -2.869, 0] \AA; $b$ = [1.658, 2.868, 0] \AA. A vacuum of 19.16 \AA ~was used along the $c$-axis to emulate a mono-layer. All states upto an energy cut-off of 70 eV above the Fermi energy were used. A {\bf k}-point grid of 20x20x1 was used. For the time propagation algorithm, a time step of .0012 fs and a total time of 120 fs was used (for details of time-propagation algorithm see Ref.~\cite{dewhurst2016}). The parameters for the Proca equation (Eq. \ref{e:proca}) were determined\cite{sharma2024direct} to be a$_0$ = 0.2 and a$_2$ = 100. The parameters used for the off-resonant (and resonant) pulse are as follows: Fluence = 1.7mJ/cm$^2$; duration = 23.95 fs; frequency of 2.12 eV (1.67 eV). A scissors correction of 0.585 eV was used for correcting the KS band-gap to experiments\cite{zhang2015probing}. 

\section{Acknowledgements} DG, Sharma, and JKD would like to thank the DFG for funding through project-ID 328545488 TRR227 (project A04). Sharma and Shallcross would like to thank Leibniz Professorin Program (SAW P118/2021) for funding. WC would like to thak DFG for funding though SH 498-8/1.

\section{Data availability}
All data involved in the production of the manuscript available upon reasonable request.

\section{Code availability}
The Elk code is freely available under GNU General Public License at https://elk.sourceforge.io/.


\begin{mcitethebibliography}{44}
\providecommand*\natexlab[1]{#1}
\providecommand*\mciteSetBstSublistMode[1]{}
\providecommand*\mciteSetBstMaxWidthForm[2]{}
\providecommand*\mciteBstWouldAddEndPuncttrue
  {\def\EndOfBibitem{\unskip.}}
\providecommand*\mciteBstWouldAddEndPunctfalse
  {\let\EndOfBibitem\relax}
\providecommand*\mciteSetBstMidEndSepPunct[3]{}
\providecommand*\mciteSetBstSublistLabelBeginEnd[3]{}
\providecommand*\EndOfBibitem{}
\mciteSetBstSublistMode{f}
\mciteSetBstMaxWidthForm{subitem}{(\alph{mcitesubitemcount})}
\mciteSetBstSublistLabelBeginEnd
  {\mcitemaxwidthsubitemform\space}
  {\relax}
  {\relax}

\bibitem[Nie \latin{et~al.}(2014)Nie, Long, Sun, Huang, Zhang, Xiong, Hewak,
  Shen, Prezhdo, and Loh]{nie2014}
Nie,~Z.; Long,~R.; Sun,~L.; Huang,~C.-C.; Zhang,~J.; Xiong,~Q.; Hewak,~D.~W.;
  Shen,~Z.; Prezhdo,~O.~V.; Loh,~Z.-H. Ultrafast carrier thermalization and
  cooling dynamics in few-layer MoS2. \emph{ACS nano} \textbf{2014}, \emph{8},
  10931--10940\relax
\mciteBstWouldAddEndPuncttrue
\mciteSetBstMidEndSepPunct{\mcitedefaultmidpunct}
{\mcitedefaultendpunct}{\mcitedefaultseppunct}\relax
\EndOfBibitem
\bibitem[Sie \latin{et~al.}(2017)Sie, Steinhoff, Gies, Lui, Ma, Rosner,
  Schönhoff, Jahnke, Wehling, Lee, \latin{et~al.} others]{sie2017}
Sie,~E.~J.; Steinhoff,~A.; Gies,~C.; Lui,~C.~H.; Ma,~Q.; Rosner,~M.;
  Schönhoff,~G.; Jahnke,~F.; Wehling,~T.~O.; Lee,~Y.-H., \latin{et~al.}
  Observation of exciton redshift--blueshift crossover in monolayer WS2.
  \emph{Nano letters} \textbf{2017}, \emph{17}, 4210--4216\relax
\mciteBstWouldAddEndPuncttrue
\mciteSetBstMidEndSepPunct{\mcitedefaultmidpunct}
{\mcitedefaultendpunct}{\mcitedefaultseppunct}\relax
\EndOfBibitem
\bibitem[Mai \latin{et~al.}(2014)Mai, Barrette, Yu, Semenov, Kim, Cao, and
  Gundogdu]{mai2014}
Mai,~C.; Barrette,~A.; Yu,~Y.; Semenov,~Y.~G.; Kim,~K.~W.; Cao,~L.;
  Gundogdu,~K. Many-body effects in valleytronics: direct measurement of valley
  lifetimes in single-layer MoS2. \emph{Nano letters} \textbf{2014}, \emph{14},
  202--206\relax
\mciteBstWouldAddEndPuncttrue
\mciteSetBstMidEndSepPunct{\mcitedefaultmidpunct}
{\mcitedefaultendpunct}{\mcitedefaultseppunct}\relax
\EndOfBibitem
\bibitem[Pogna \latin{et~al.}(2016)Pogna, Marsili, De~Fazio, Dal~Conte,
  Manzoni, Sangalli, Yoon, Lombardo, Ferrari, Marini, \latin{et~al.}
  others]{pogna2016}
Pogna,~E.~A.; Marsili,~M.; De~Fazio,~D.; Dal~Conte,~S.; Manzoni,~C.;
  Sangalli,~D.; Yoon,~D.; Lombardo,~A.; Ferrari,~A.~C.; Marini,~A.,
  \latin{et~al.}  Photo-induced bandgap renormalization governs the ultrafast
  response of single-layer MoS2. \emph{ACS nano} \textbf{2016}, \emph{10},
  1182--1188\relax
\mciteBstWouldAddEndPuncttrue
\mciteSetBstMidEndSepPunct{\mcitedefaultmidpunct}
{\mcitedefaultendpunct}{\mcitedefaultseppunct}\relax
\EndOfBibitem
\bibitem[Shang \latin{et~al.}(2015)Shang, Shen, Cong, Peimyoo, Cao, Eginligil,
  and Yu]{shang2015}
Shang,~J.; Shen,~X.; Cong,~C.; Peimyoo,~N.; Cao,~B.; Eginligil,~M.; Yu,~T.
  Observation of excitonic fine structure in a 2D transition-metal
  dichalcogenide semiconductor. \emph{ACS nano} \textbf{2015}, \emph{9},
  647--655\relax
\mciteBstWouldAddEndPuncttrue
\mciteSetBstMidEndSepPunct{\mcitedefaultmidpunct}
{\mcitedefaultendpunct}{\mcitedefaultseppunct}\relax
\EndOfBibitem
\bibitem[Lucchini \latin{et~al.}(2021)Lucchini, Sato, Lucarelli, Moio, Inzani,
  Borrego-Varillas, Frassetto, Poletto, H{\"u}bener, De~Giovannini,
  \latin{et~al.} others]{lucchini2021}
Lucchini,~M.; Sato,~S.~A.; Lucarelli,~G.~D.; Moio,~B.; Inzani,~G.;
  Borrego-Varillas,~R.; Frassetto,~F.; Poletto,~L.; H{\"u}bener,~H.;
  De~Giovannini,~U., \latin{et~al.}  Unravelling the intertwined atomic and
  bulk nature of localised excitons by attosecond spectroscopy. \emph{Nature
  communications} \textbf{2021}, \emph{12}, 1021\relax
\mciteBstWouldAddEndPuncttrue
\mciteSetBstMidEndSepPunct{\mcitedefaultmidpunct}
{\mcitedefaultendpunct}{\mcitedefaultseppunct}\relax
\EndOfBibitem
\bibitem[Sim \latin{et~al.}(2013)Sim, Park, Song, In, Lee, Kim, and
  Choi]{sim2013}
Sim,~S.; Park,~J.; Song,~J.-G.; In,~C.; Lee,~Y.-S.; Kim,~H.; Choi,~H. Exciton
  dynamics in atomically thin MoS 2: interexcitonic interaction and broadening
  kinetics. \emph{Physical Review B} \textbf{2013}, \emph{88}, 075434\relax
\mciteBstWouldAddEndPuncttrue
\mciteSetBstMidEndSepPunct{\mcitedefaultmidpunct}
{\mcitedefaultendpunct}{\mcitedefaultseppunct}\relax
\EndOfBibitem
\bibitem[Kirilyuk \latin{et~al.}(2010)Kirilyuk, Kimel, and
  Rasing]{kirilyuk2010ultrafast}
Kirilyuk,~A.; Kimel,~A.~V.; Rasing,~T. Ultrafast optical manipulation of
  magnetic order. \emph{Reviews of Modern Physics} \textbf{2010}, \emph{82},
  2731--2784\relax
\mciteBstWouldAddEndPuncttrue
\mciteSetBstMidEndSepPunct{\mcitedefaultmidpunct}
{\mcitedefaultendpunct}{\mcitedefaultseppunct}\relax
\EndOfBibitem
\bibitem[Gon{\c{c}}alves \latin{et~al.}(2016)Gon{\c{c}}alves, Silva, Navas,
  Miranda, Silva, Crespo, and Schmool]{gonccalves2016dual}
Gon{\c{c}}alves,~C.; Silva,~A.; Navas,~D.; Miranda,~M.; Silva,~F.; Crespo,~H.;
  Schmool,~D. A dual-colour architecture for pump-probe spectroscopy of
  ultrafast magnetization dynamics in the sub-10-femt osecond range.
  \emph{Scientific reports} \textbf{2016}, \emph{6}, 22872\relax
\mciteBstWouldAddEndPuncttrue
\mciteSetBstMidEndSepPunct{\mcitedefaultmidpunct}
{\mcitedefaultendpunct}{\mcitedefaultseppunct}\relax
\EndOfBibitem
\bibitem[Zuti{\'c} \latin{et~al.}(2004)Zuti{\'c}, Fabian, and
  Sarma]{vzutic2004}
Zuti{\'c},~I.; Fabian,~J.; Sarma,~S.~D. Spintronics: Fundamentals and
  applications. \emph{Reviews of modern physics} \textbf{2004}, \emph{76},
  323\relax
\mciteBstWouldAddEndPuncttrue
\mciteSetBstMidEndSepPunct{\mcitedefaultmidpunct}
{\mcitedefaultendpunct}{\mcitedefaultseppunct}\relax
\EndOfBibitem
\bibitem[Ouyang \latin{et~al.}(2020)Ouyang, Chen, Tang, Zhang, Zhang, Zhang,
  Cheng, and Jiang]{ouyang2020}
Ouyang,~H.; Chen,~H.; Tang,~Y.; Zhang,~J.; Zhang,~C.; Zhang,~B.; Cheng,~X.;
  Jiang,~T. All-optical dynamic tuning of local excitonic emission of monolayer
  MoS2 by integration with Ge2Sb2Te5. \emph{Nanophotonics} \textbf{2020},
  \emph{9}, 2351--2359\relax
\mciteBstWouldAddEndPuncttrue
\mciteSetBstMidEndSepPunct{\mcitedefaultmidpunct}
{\mcitedefaultendpunct}{\mcitedefaultseppunct}\relax
\EndOfBibitem
\bibitem[Kobayashi \latin{et~al.}(2023)Kobayashi, Heide, Johnson, Tiwari, Liu,
  Reis, Heinz, and Ghimire]{kobayashi2023}
Kobayashi,~Y.; Heide,~C.; Johnson,~A.~C.; Tiwari,~V.; Liu,~F.; Reis,~D.~A.;
  Heinz,~T.~F.; Ghimire,~S. Floquet engineering of strongly driven excitons in
  monolayer tungsten disulfide. \emph{Nature Physics} \textbf{2023}, \emph{19},
  171--176\relax
\mciteBstWouldAddEndPuncttrue
\mciteSetBstMidEndSepPunct{\mcitedefaultmidpunct}
{\mcitedefaultendpunct}{\mcitedefaultseppunct}\relax
\EndOfBibitem
\bibitem[Krieger \latin{et~al.}(2015)Krieger, Dewhurst, Elliott, Sharma, and
  Gross]{krieger2015laser}
Krieger,~K.; Dewhurst,~J.; Elliott,~P.; Sharma,~S.; Gross,~E. Laser-induced
  demagnetization at ultrashort time scales: Predictions of TDDFT.
  \emph{Journal of chemical theory and computation} \textbf{2015}, \emph{11},
  4870--4874\relax
\mciteBstWouldAddEndPuncttrue
\mciteSetBstMidEndSepPunct{\mcitedefaultmidpunct}
{\mcitedefaultendpunct}{\mcitedefaultseppunct}\relax
\EndOfBibitem
\bibitem[Liao \latin{et~al.}(2023)Liao, Kasper, Jenkins, Yang, Batista, Frisch,
  and Li]{liao2023state}
Liao,~C.; Kasper,~J.~M.; Jenkins,~A.~J.; Yang,~P.; Batista,~E.~R.;
  Frisch,~M.~J.; Li,~X. State Interaction Linear Response Time-Dependent
  Density Functional Theory with Perturbative Spin--Orbit Coupling: Benchmark
  and Perspectives. \emph{JACS Au} \textbf{2023}, \emph{3}, 358--367\relax
\mciteBstWouldAddEndPuncttrue
\mciteSetBstMidEndSepPunct{\mcitedefaultmidpunct}
{\mcitedefaultendpunct}{\mcitedefaultseppunct}\relax
\EndOfBibitem
\bibitem[L{\'o}pez \latin{et~al.}(2013)L{\'o}pez, Scholz, Sun, and
  Schliemann]{lopez2013graphene}
L{\'o}pez,~A.; Scholz,~A.; Sun,~Z.; Schliemann,~J. Graphene with time-dependent
  spin-orbit coupling: truncated Magnus expansion approach. \emph{The European
  Physical Journal B} \textbf{2013}, \emph{86}, 1--8\relax
\mciteBstWouldAddEndPuncttrue
\mciteSetBstMidEndSepPunct{\mcitedefaultmidpunct}
{\mcitedefaultendpunct}{\mcitedefaultseppunct}\relax
\EndOfBibitem
\bibitem[Dewhurst \latin{et~al.}(2025)Dewhurst, Gill, Shallcross, and
  Sharma]{dewhurst2025kohn}
Dewhurst,~J.; Gill,~D.; Shallcross,~S.; Sharma,~S. Kohn-Sham-Proca equations
  for ultrafast exciton dynamics. \emph{Physical Review B} \textbf{2025},
  \emph{111}, L060302\relax
\mciteBstWouldAddEndPuncttrue
\mciteSetBstMidEndSepPunct{\mcitedefaultmidpunct}
{\mcitedefaultendpunct}{\mcitedefaultseppunct}\relax
\EndOfBibitem
\bibitem[Runge and Gross(1984)Runge, and Gross]{runge1984}
Runge,~E.; Gross,~E.~K. Density-functional theory for time-dependent systems.
  \emph{Physical review letters} \textbf{1984}, \emph{52}, 997\relax
\mciteBstWouldAddEndPuncttrue
\mciteSetBstMidEndSepPunct{\mcitedefaultmidpunct}
{\mcitedefaultendpunct}{\mcitedefaultseppunct}\relax
\EndOfBibitem
\bibitem[Aslan \latin{et~al.}(2018)Aslan, Deng, and Heinz]{aslan2018strain}
Aslan,~B.; Deng,~M.; Heinz,~T.~F. Strain tuning of excitons in monolayer WSe 2.
  \emph{Physical Review B} \textbf{2018}, \emph{98}, 115308\relax
\mciteBstWouldAddEndPuncttrue
\mciteSetBstMidEndSepPunct{\mcitedefaultmidpunct}
{\mcitedefaultendpunct}{\mcitedefaultseppunct}\relax
\EndOfBibitem
\bibitem[Zhang \latin{et~al.}(2015)Zhang, Chen, Johnson, Li, Li, Mende,
  Feenstra, and Shih]{zhang2015probing}
Zhang,~C.; Chen,~Y.; Johnson,~A.; Li,~M.-Y.; Li,~L.-J.; Mende,~P.~C.;
  Feenstra,~R.~M.; Shih,~C.-K. Probing critical point energies of transition
  metal dichalcogenides: surprising indirect gap of single layer WSe2.
  \emph{Nano letters} \textbf{2015}, \emph{15}, 6494--6500\relax
\mciteBstWouldAddEndPuncttrue
\mciteSetBstMidEndSepPunct{\mcitedefaultmidpunct}
{\mcitedefaultendpunct}{\mcitedefaultseppunct}\relax
\EndOfBibitem
\bibitem[Aslan \latin{et~al.}(2021)Aslan, Yule, Yu, Lee, Heinz, Cao, and
  Brongersma]{aslan2021excitons}
Aslan,~B.; Yule,~C.; Yu,~Y.; Lee,~Y.~J.; Heinz,~T.~F.; Cao,~L.;
  Brongersma,~M.~L. Excitons in strained and suspended monolayer WSe2. \emph{2D
  Materials} \textbf{2021}, \emph{9}, 015002\relax
\mciteBstWouldAddEndPuncttrue
\mciteSetBstMidEndSepPunct{\mcitedefaultmidpunct}
{\mcitedefaultendpunct}{\mcitedefaultseppunct}\relax
\EndOfBibitem
\bibitem[Sharma \latin{et~al.}(2024)Sharma, Gill, Krishna, Dewhurst, and
  Shallcross]{sharma2024direct}
Sharma,~S.; Gill,~D.; Krishna,~J.; Dewhurst,~J.; Shallcross,~S. Direct coupling
  of light to valley current. \emph{Nature Communications} \textbf{2024},
  \emph{15}, 7579\relax
\mciteBstWouldAddEndPuncttrue
\mciteSetBstMidEndSepPunct{\mcitedefaultmidpunct}
{\mcitedefaultendpunct}{\mcitedefaultseppunct}\relax
\EndOfBibitem
\bibitem[Aharon \latin{et~al.}(2016)Aharon, Gdor, Yang, Etgar, Ruhman,
  \latin{et~al.} others]{aharon2016new}
Aharon,~S.; Gdor,~I.; Yang,~C.; Etgar,~L.; Ruhman,~S., \latin{et~al.}  New
  insights into exciton binding and relaxation from high time resolution
  ultrafast spectroscopy of CH 3 NH 3 PbI 3 and CH 3 NH 3 PbBr 3 films.
  \emph{Journal of Materials Chemistry A} \textbf{2016}, \emph{4},
  3546--3553\relax
\mciteBstWouldAddEndPuncttrue
\mciteSetBstMidEndSepPunct{\mcitedefaultmidpunct}
{\mcitedefaultendpunct}{\mcitedefaultseppunct}\relax
\EndOfBibitem
\bibitem[Baldini \latin{et~al.}(2018)Baldini, Palmieri, Pomarico, Aubock, and
  Chergui]{baldini2018clocking}
Baldini,~E.; Palmieri,~T.; Pomarico,~E.; Aubock,~G.; Chergui,~M. Clocking the
  ultrafast electron cooling in anatase titanium dioxide nanoparticles.
  \emph{ACS Photonics} \textbf{2018}, \emph{5}, 1241--1249\relax
\mciteBstWouldAddEndPuncttrue
\mciteSetBstMidEndSepPunct{\mcitedefaultmidpunct}
{\mcitedefaultendpunct}{\mcitedefaultseppunct}\relax
\EndOfBibitem
\bibitem[Fluegel \latin{et~al.}(1997)Fluegel, Zhang, Cheong, Mascarenhas,
  Geisz, Olson, and Duda]{fluegel1997exciton}
Fluegel,~B.; Zhang,~Y.; Cheong,~H.; Mascarenhas,~A.; Geisz,~J.; Olson,~J.;
  Duda,~A. Exciton absorption bleaching studies in ordered Ga x In 1- x P.
  \emph{Physical Review B} \textbf{1997}, \emph{55}, 13647\relax
\mciteBstWouldAddEndPuncttrue
\mciteSetBstMidEndSepPunct{\mcitedefaultmidpunct}
{\mcitedefaultendpunct}{\mcitedefaultseppunct}\relax
\EndOfBibitem
\bibitem[Timmer \latin{et~al.}(2024)Timmer, Gittinger, Quenzel, Cadore, Rosa,
  Li, Soavi, Lunemann, Stephan, Silies, Tommy, \latin{et~al.}
  others]{timmer2024ultrafast}
Timmer,~D.; Gittinger,~M.; Quenzel,~T.; Cadore,~A.~R.; Rosa,~B.~L.; Li,~W.;
  Soavi,~G.; Lunemann,~D.~C.; Stephan,~S.; Silies,~M.; Tommy,~S.,
  \latin{et~al.}  Ultrafast coherent exciton couplings and many-body
  interactions in monolayer WS2. \emph{Nano Letters} \textbf{2024}, \emph{24},
  8117--8125\relax
\mciteBstWouldAddEndPuncttrue
\mciteSetBstMidEndSepPunct{\mcitedefaultmidpunct}
{\mcitedefaultendpunct}{\mcitedefaultseppunct}\relax
\EndOfBibitem
\bibitem[Souri \latin{et~al.}(2024)Souri, Timmer, Lunemann, Hadilou, Winte,
  De~Sio, Esmann, Curdt, Winklhofer, Anhauser, \latin{et~al.}
  others]{souri2024ultrafast}
Souri,~S.; Timmer,~D.; Lunemann,~D.~C.; Hadilou,~N.; Winte,~K.; De~Sio,~A.;
  Esmann,~M.; Curdt,~F.; Winklhofer,~M.; Anhauser,~S., \latin{et~al.}
  Ultrafast Time-Domain Spectroscopy Reveals Coherent Vibronic Couplings upon
  Electronic Excitation in Crystalline Organic Thin Films. \emph{The Journal of
  Physical Chemistry Letters} \textbf{2024}, \emph{15}, 11170--11181\relax
\mciteBstWouldAddEndPuncttrue
\mciteSetBstMidEndSepPunct{\mcitedefaultmidpunct}
{\mcitedefaultendpunct}{\mcitedefaultseppunct}\relax
\EndOfBibitem
\bibitem[Dong \latin{et~al.}(2021)Dong, Puppin, Pincelli, Beaulieu,
  Christiansen, H{\"u}bener, Nicholson, Xian, Dendzik, Deng, \latin{et~al.}
  others]{dong2021direct}
Dong,~S.; Puppin,~M.; Pincelli,~T.; Beaulieu,~S.; Christiansen,~D.;
  H{\"u}bener,~H.; Nicholson,~C.~W.; Xian,~R.~P.; Dendzik,~M.; Deng,~Y.,
  \latin{et~al.}  Direct measurement of key exciton properties: Energy,
  dynamics, and spatial distribution of the wave function. \emph{Natural
  sciences} \textbf{2021}, \emph{1}, e10010\relax
\mciteBstWouldAddEndPuncttrue
\mciteSetBstMidEndSepPunct{\mcitedefaultmidpunct}
{\mcitedefaultendpunct}{\mcitedefaultseppunct}\relax
\EndOfBibitem
\bibitem[Ma \latin{et~al.}(2005)Ma, Valkunas, Dexheimer, Bachilo, and
  Fleming]{ma2005femtosecond}
Ma,~Y.-Z.; Valkunas,~L.; Dexheimer,~S.~L.; Bachilo,~S.~M.; Fleming,~G.~R.
  Femtosecond Spectroscopy of Optical Excitations in Single-Walled Carbon
  Nanotubes:<? format?> Evidence for Exciton-Exciton Annihilation.
  \emph{Physical review letters} \textbf{2005}, \emph{94}, 157402\relax
\mciteBstWouldAddEndPuncttrue
\mciteSetBstMidEndSepPunct{\mcitedefaultmidpunct}
{\mcitedefaultendpunct}{\mcitedefaultseppunct}\relax
\EndOfBibitem
\bibitem[Smejkal \latin{et~al.}(2020)Smejkal, Libisch, Molina-Sanchez,
  Trovatello, Wirtz, and Marini]{smejkal2020}
Smejkal,~V.; Libisch,~F.; Molina-Sanchez,~A.; Trovatello,~C.; Wirtz,~L.;
  Marini,~A. Time-dependent screening explains the ultrafast excitonic signal
  rise in 2D semiconductors. \emph{ACS nano} \textbf{2020}, \emph{15},
  1179--1185\relax
\mciteBstWouldAddEndPuncttrue
\mciteSetBstMidEndSepPunct{\mcitedefaultmidpunct}
{\mcitedefaultendpunct}{\mcitedefaultseppunct}\relax
\EndOfBibitem
\bibitem[Cunningham \latin{et~al.}(2017)Cunningham, Hanbicki, McCreary, and
  Jonker]{cunningham2017photoinduced}
Cunningham,~P.~D.; Hanbicki,~A.~T.; McCreary,~K.~M.; Jonker,~B.~T. Photoinduced
  bandgap renormalization and exciton binding energy reduction in WS2.
  \emph{ACS nano} \textbf{2017}, \emph{11}, 12601--12608\relax
\mciteBstWouldAddEndPuncttrue
\mciteSetBstMidEndSepPunct{\mcitedefaultmidpunct}
{\mcitedefaultendpunct}{\mcitedefaultseppunct}\relax
\EndOfBibitem
\bibitem[Chernikov \latin{et~al.}(2015)Chernikov, Ruppert, Hill, Rigosi, and
  Heinz]{chernikov2015population}
Chernikov,~A.; Ruppert,~C.; Hill,~H.~M.; Rigosi,~A.~F.; Heinz,~T.~F. Population
  inversion and giant bandgap renormalization in atomically thin WS2 layers.
  \emph{Nature Photonics} \textbf{2015}, \emph{9}, 466--470\relax
\mciteBstWouldAddEndPuncttrue
\mciteSetBstMidEndSepPunct{\mcitedefaultmidpunct}
{\mcitedefaultendpunct}{\mcitedefaultseppunct}\relax
\EndOfBibitem
\bibitem[Liu \latin{et~al.}(2019)Liu, Ziffer, Hansen, Wang, and
  Zhu]{liu2019direct}
Liu,~F.; Ziffer,~M.~E.; Hansen,~K.~R.; Wang,~J.; Zhu,~X. Direct determination
  of band-gap renormalization in the photoexcited monolayer MoS 2.
  \emph{Physical review letters} \textbf{2019}, \emph{122}, 246803\relax
\mciteBstWouldAddEndPuncttrue
\mciteSetBstMidEndSepPunct{\mcitedefaultmidpunct}
{\mcitedefaultendpunct}{\mcitedefaultseppunct}\relax
\EndOfBibitem
\bibitem[Zhu \latin{et~al.}(2018)Zhu, Liu, Wang, Peng, Qi, Bao, Fan, and
  Wang]{zhu2018breaking}
Zhu,~Y.; Liu,~L.; Wang,~J.; Peng,~R.; Qi,~D.; Bao,~W.; Fan,~R.; Wang,~M.
  Breaking Pauli blockade via ultrafast cooling of hot electrons in
  optically-pumped graphene. \emph{arXiv preprint arXiv:1806.04453}
  \textbf{2018}, \relax
\mciteBstWouldAddEndPunctfalse
\mciteSetBstMidEndSepPunct{\mcitedefaultmidpunct}
{}{\mcitedefaultseppunct}\relax
\EndOfBibitem
\bibitem[Iwasaki \latin{et~al.}(2023)Iwasaki, Fukuda, Noyama, Akei, Shigekawa,
  Fons, Hase, Arashida, and Hada]{iwasaki2023electronic}
Iwasaki,~Y.; Fukuda,~T.; Noyama,~G.; Akei,~M.; Shigekawa,~H.; Fons,~P.~J.;
  Hase,~M.; Arashida,~Y.; Hada,~M. Electronic intraband scattering in a
  transition-metal dichalcogenide observed by double-excitation ultrafast
  electron diffraction. \emph{Applied Physics Letters} \textbf{2023},
  \emph{123}\relax
\mciteBstWouldAddEndPuncttrue
\mciteSetBstMidEndSepPunct{\mcitedefaultmidpunct}
{\mcitedefaultendpunct}{\mcitedefaultseppunct}\relax
\EndOfBibitem
\bibitem[Amo \latin{et~al.}(2007)Amo, Vi{\~n}a, Lugli, Tejedor, Toropov, and
  Zhuravlev]{amo2007pauli}
Amo,~A.; Vi{\~n}a,~L.; Lugli,~P.; Tejedor,~C.; Toropov,~A.; Zhuravlev,~K. Pauli
  blockade of the electron spin flip in bulk GaAs. \emph{Physical Review
  B—Condensed Matter and Materials Physics} \textbf{2007}, \emph{75},
  085202\relax
\mciteBstWouldAddEndPuncttrue
\mciteSetBstMidEndSepPunct{\mcitedefaultmidpunct}
{\mcitedefaultendpunct}{\mcitedefaultseppunct}\relax
\EndOfBibitem
\bibitem[Sun \latin{et~al.}(2014)Sun, Rao, Reider, Chen, You, Br{\'e}zin,
  Harutyunyan, and Heinz]{sun2014observation}
Sun,~D.; Rao,~Y.; Reider,~G.~A.; Chen,~G.; You,~Y.; Br{\'e}zin,~L.;
  Harutyunyan,~A.~R.; Heinz,~T.~F. Observation of rapid exciton--exciton
  annihilation in monolayer molybdenum disulfide. \emph{Nano letters}
  \textbf{2014}, \emph{14}, 5625--5629\relax
\mciteBstWouldAddEndPuncttrue
\mciteSetBstMidEndSepPunct{\mcitedefaultmidpunct}
{\mcitedefaultendpunct}{\mcitedefaultseppunct}\relax
\EndOfBibitem
\bibitem[Wang \latin{et~al.}(2015)Wang, Zhang, and Rana]{wang2015surface}
Wang,~H.; Zhang,~C.; Rana,~F. Surface recombination limited lifetimes of
  photoexcited carriers in few-layer transition metal dichalcogenide MoS2.
  \emph{Nano letters} \textbf{2015}, \emph{15}, 8204--8210\relax
\mciteBstWouldAddEndPuncttrue
\mciteSetBstMidEndSepPunct{\mcitedefaultmidpunct}
{\mcitedefaultendpunct}{\mcitedefaultseppunct}\relax
\EndOfBibitem
\bibitem[Kumar \latin{et~al.}(2014)Kumar, Cui, Ceballos, He, Wang, and
  Zhao]{kumar2014exciton}
Kumar,~N.; Cui,~Q.; Ceballos,~F.; He,~D.; Wang,~Y.; Zhao,~H. Exciton-exciton
  annihilation in MoSe 2 monolayers. \emph{Physical Review B} \textbf{2014},
  \emph{89}, 125427\relax
\mciteBstWouldAddEndPuncttrue
\mciteSetBstMidEndSepPunct{\mcitedefaultmidpunct}
{\mcitedefaultendpunct}{\mcitedefaultseppunct}\relax
\EndOfBibitem
\bibitem[Kar \latin{et~al.}(2015)Kar, Su, Nair, and Sood]{kar2015probing}
Kar,~S.; Su,~Y.; Nair,~R.~R.; Sood,~A. Probing photoexcited carriers in a
  few-layer MoS2 laminate by time-resolved optical pump--terahertz probe
  spectroscopy. \emph{ACS nano} \textbf{2015}, \emph{9}, 12004--12010\relax
\mciteBstWouldAddEndPuncttrue
\mciteSetBstMidEndSepPunct{\mcitedefaultmidpunct}
{\mcitedefaultendpunct}{\mcitedefaultseppunct}\relax
\EndOfBibitem
\bibitem[Sharma \latin{et~al.}(2014)Sharma, Dewhurst, and Gross]{sharma2014}
Sharma,~S.; Dewhurst,~J.; Gross,~E. Optical response of extended systems using
  time-dependent density functional theory. \emph{First Principles Approaches
  to Spectroscopic Properties of Complex Materials} \textbf{2014},
  235--257\relax
\mciteBstWouldAddEndPuncttrue
\mciteSetBstMidEndSepPunct{\mcitedefaultmidpunct}
{\mcitedefaultendpunct}{\mcitedefaultseppunct}\relax
\EndOfBibitem
\bibitem[Dewhurst \latin{et~al.}(Jan. 14 {\bf 2018})Dewhurst, Sharma, and
  et~al.]{elk}
Dewhurst,~J.~K.; Sharma,~S.; et~al., Jan. 14 {\bf 2018};
  \url{elk.sourceforge.net}\relax
\mciteBstWouldAddEndPuncttrue
\mciteSetBstMidEndSepPunct{\mcitedefaultmidpunct}
{\mcitedefaultendpunct}{\mcitedefaultseppunct}\relax
\EndOfBibitem
\bibitem[Schutte \latin{et~al.}(1987)Schutte, De~Boer, and
  Jellinek]{schutte1987crystal}
Schutte,~W.; De~Boer,~J.; Jellinek,~F. Crystal structures of tungsten disulfide
  and diselenide. \emph{Journal of Solid State Chemistry} \textbf{1987},
  \emph{70}, 207--209\relax
\mciteBstWouldAddEndPuncttrue
\mciteSetBstMidEndSepPunct{\mcitedefaultmidpunct}
{\mcitedefaultendpunct}{\mcitedefaultseppunct}\relax
\EndOfBibitem
\bibitem[Dewhurst \latin{et~al.}(2016)Dewhurst, Krieger, Sharma, and
  Gross]{dewhurst2016}
Dewhurst,~J.~K.; Krieger,~K.; Sharma,~S.; Gross,~E. K.~U. An efficient
  algorithm for time propagation as applied to linearized augmented plane wave
  method. \emph{Computer Physics Communications} \textbf{2016}, \emph{209},
  92--95\relax
\mciteBstWouldAddEndPuncttrue
\mciteSetBstMidEndSepPunct{\mcitedefaultmidpunct}
{\mcitedefaultendpunct}{\mcitedefaultseppunct}\relax
\EndOfBibitem
\end{mcitethebibliography}
\providecommand{\latin}[1]{#1}
\makeatletter
\providecommand{\doi}
  {\begingroup\let\do\@makeother\dospecials
  \catcode`\{=1 \catcode`\}=2 \doi@aux}
\providecommand{\doi@aux}[1]{\endgroup\texttt{#1}}
\makeatother
\providecommand*\mcitethebibliography{\thebibliography}
\csname @ifundefined\endcsname{endmcitethebibliography}
  {\let\endmcitethebibliography\endthebibliography}{}

\end{document}